\documentstyle[preprint,aps,pra]{revtex}
\begin{document}
\newcommand\beq{\begin{equation}}
\newcommand\eeq{\end{equation}}
\newcommand\bea{\begin{eqnarray}}
\newcommand\eea{\end{eqnarray}} 

\newcommand\gt{\tilde{g}}
\tightenlines
\title{Number Fluctuation in an interacting trapped gas in one and two 
dimensions}
\author{R. K. Bhaduri$^1$ , M. V. N. Murthy and Muoi N. Tran$^1$} 
\address{ The Institute of Mathematical Sciences\\ 
Chennai 600 113, India\\
1. Permanent Address: Department of Physics and Astronomy,\\
McMaster University,Hamilton, Ontario,\\ Canada L8S 4M1
}
\maketitle

\begin{abstract}
It is well-known that the number fluctuation in the grand canonical ensemble, 
which is directly proportional to the compressibility, diverges for an ideal   
bose gas as $T\rightarrow 0$. 
We show that this divergence is
removed when the atoms interact in one dimension through an inverse square
two-body interaction. In two dimensions, similar results are obtained
using a self-consistent Thomas-Fermi (TF) model for a repulsive zero-range
interaction. Both models may be mapped on to a system of non-interacting
particles obeying the Haldane-Wu exclusion statistics. We also calculate
the number fluctuation from the ground state of the gas in these
interacting models, and compare the grand canonical results with those 
obtained from the canonical ensemble. 
\vskip .5 true cm
\end{abstract}

\section{Introduction}

Consider an ideal bose gas at low temperatures. In the grand canonical
ensemble (GCE), the compressibility and the number fluctuation of this system
diverge at low temperatures~\cite{landau}.  To quote Landau and 
Lifshitz~\cite{landau}:  
\begin{quote}

...in a bose gas at temperature $T < T_c$, the pressure is independent of
the volume, i.e. the compressibility becomes infinite.  Accordingly...this
would imply that the fluctuations of the number of particles also become
infinite. This means that, in calculating fluctuations in a gas obeying
bose statistics, {\it the interactions between the particles cannot be
neglected at low temperatures, however weak this interaction may be. When
the interactions, which must exist in any actual gas, is taken into
account, the resulting fluctuations are finite}. 
\end{quote}

One purpose of this paper is to demonstrate the validity of this statement
in one and two dimensional traps for a special class of interactions. 

In one dimension, the quantum many-body problem of particles in a harmonic
oscillator interacting with an inverse square two-body potential is
exactly solvable~\cite{calogero,sutherland}. Moreover, it is known that
the global properties of these interacting bosons are the same as those of
non-interacting particles obeying the Haldane-Wu generalized exclusion
statistics (ideal haldons)~\cite{haldane,murthy}.  Using this mapping, we
show that the number fluctuation is finite as $T\rightarrow 0$, no matter
how weak the interaction strength is.  In this model, since the exact
correlation function is known for some specific strength parameters of the
interaction~\cite{sutherland}, it is also possible to verify explicitly
the well-known relation between its integral and the number fluctuation. 

In two dimensions, there is no suitable exactly solvable model for our
purpose. We therefore consider the mean-field model of bosons trapped in a
harmonic oscillator, and interacting pair-wise with a zero-range repulsive
pseudo-potential~\cite{bhaduri}. In the absence of this interaction, the
number fluctuation diverges at $T=T_c$. When the interaction is present,
however weak, we show that the compressibility and hence the number
fluctuation of the system is finite right down to $T=0$. Moreover, they
are shown to be identical to those of a collection of non-interacting
haldons. 
 
Another quantity of some interest is the number fluctuation of particles
from the {\it ground state} of the system, which is present even when the
total number of particles in the trap is fixed. Consider a dilute gas of
bosons in a trap at $T=0$.  The system is in its ground state. When a
certain amount of excitation energy is given to the system, it may be
absorbed in many possible ways, so that the number of bosons remaining in
the ground state is not fixed.  This number fluctuation for non-interacting
bosons in a harmonic trap has been calculated by a number of
authors~\cite{grossman} as a function of the excitation energy or
temperature. Thermodynamic identities and particle number fluctuations in
weakly interacting BEC have also been analysed when the particle number is
fixed\cite{illuminati}.  For ideal bosons, the number fluctuation from the
ground state diverges at low temperatures in GCE, but this can be avoided
using more careful canonical, or microcanonical treatment. 
When the inverse square pair-wise
interaction is used in one dimension, this divergence is again removed.
Further, in this interacting model, we can also perform the canonical
ensemble (CE) calculations, and compare with the GCE results. We find that
even though the ground state number fluctuation goes to zero as $T
\rightarrow 0$ in both GCE and CE, the very low-temperature behaviors are
rather different . At higher temperatures the results from CE and GCE tend
to coincide. Similar studies can be made for fermionic systems at low
temperatures. Since the interactions that we have used are of statistical
character~\cite{murthy,hansson}, our results interpolate from bosonic to
fermionic behavior with the variation in the strength of the interaction. 

The outline of the paper is as follows: Sect.~II deals with the total
number fluctuation in GCE for interacting models.  Sect.~II (a) is devoted
to a discussion of the fluctuations in one dimension, and II (b) to
fluctuation in two dimensions.  In Sect.~III, the ground state number
fluctuations for the one-dimensional interacting model in GCE and CE are
calculated and compared. Particular attention is paid to the low
temperature behavior (see also  Appendix). We conclude 
the paper with a brief discussion of the results. 

\section{Fluctuations in GCE}

The number fluctuation in a gas  
in the GCE formalism is defined by $(\delta N)^2 = (<N^2>-<N>^2)$, where 
the angular brackets denote ensemble averaging. For an ideal bose gas,  
\beq
(\delta N)^2  =  \sum_{k=0}^{\infty} \langle  n_k \rangle (\langle  
n_k \rangle +1),
\label{eq1}
\eeq 
where the single particle occupancy $\langle n_k \rangle$ at a 
given temperature $T$ and energy $\epsilon_k$ are given by the bose 
distribution function 
\beq
\langle  n_k \rangle = \frac{1}{\exp[\beta(\epsilon_k - \mu)] -1}.
\eeq
Here $\beta = 1/T$ with the Boltzmann constant $k_B=1$, and $\mu$ denotes
the chemical potential. The isothermal compressibility $\chi_{_T}$ of a gas 
of density $\rho_0$ is 
related to the number fluctuation of the system in GCE :
\beq
{(\delta N)^2\over N} = T \rho_{0} \chi_{_T}~.
\label{compressibility}
\eeq 

The problem with the number fluctuation in a bose
gas in GCE is obvious from Eq.~(\ref{eq1}). If there is BEC, then a
macroscopic fraction of the particles occupy the ground state for $T<T_c$,
so that $(\delta N_0)^2\simeq N^2$. Even if there is no BEC, the same is
the case as $T\rightarrow 0$. In the thermodynamic limit, therefore, the
fluctuation diverges below the critical temperature, or in any case at
$T=0$. This is manifestly incorrect, since at $T=0$, all the bosons are in
the ground state, and the number fluctuation should vanish. In the
presence of interactions, however, this number fluctuation is expected to
be finite.  We demonstrate this by analyzing two models in one and two
dimensions. 

\subsection{Fluctuations in a one dimensional model}

We first consider the exactly solvable one dimensional 
Calogero-Sutherland Model (CSM) of a system of interacting particles with 
the Hamiltonian \cite{calogero,sutherland}    
\begin{equation}
H = \sum_{i=1}^N \left [ -\frac{\hbar^2}{2m} \frac{\partial^2}{\partial 
x_i^2} + \frac{1}{2} m\omega^2 x_i^2 \right ] +
\frac{\hbar^2}{m} \sum_{i<j=1}^N \frac{g(g-1)}{(x_i-x_j)^2},
\label{ham}
\end{equation}
with the dimensionless coupling parameter $g \geq 0$. The particles are 
confined in a harmonic well and the thermodynamic limit
is obtained by taking $\omega \rightarrow 0$ as $N\rightarrow \infty$, with 
$\omega N=constant$. In the thermodynamic limit, the properties of the system 
are translationally invariant, and would be the same if the particles were 
on a line, or a circle, instead of a harmonic confinement. To make the 
problem well-defined quantum mechanically, we have to demand that the wave functions go to zero as $|x_i-x_j|^g$ whenever two particles i and j approach 
each other. Since the particles cannot cross each other, we may choose the 
wave function to be either symmetric (bosonic) or antisymmetric (fermionic). 
For $g=0$ and $1$, the model describes free bosons and free 
fermions respectively.  

Using the relation between the integral of the correlation function and
the number fluctuation, we now show that the number fluctuation vanishes
at zero temperature in the above interacting model unlike the ideal boson
result in GCE. If $\nu(r)$ denotes the two-particle ground 
state density-density correlation function in the ground state, with  
$r=|x_1-x_2|$, then the number fluctuation is related to the correlation 
function as  \cite{landau} 
\beq
\frac{(\delta N)^2}{N} -1  =  - \int_{-\infty}^{\infty} \nu(x)dx.
\label{eq2}
\eeq 
Note that the ground state correlation function $\nu(r)$ is defined only
for $r\ge 0$. However, in computing the above integral it is necessary to
assume $\nu(r)$ to be even function, and extend the domain of integration
to negative values of $r$\cite{mehta}. Unlike the one-particle off-diagonal
density matrix, $\nu(r)$, by definition, is related to the diagonal
element of the the two-particle density matrix, and is the same in CSM for
bosons or fermions. In this section we work in the bosonic basis.  The
correlation functions are known exactly in the CSM for three values of $g$
independent of whether the particles are bosons are fermions and are given
(in the thermodynamic limit) by \cite{sutherland}
\bea
g&=&1:~~ \nu(r) = s(r)^2 = \left[ \frac{\sin(\pi r)}{\pi r}\right]^2\\
g&=&1/2:\nu(r) = s(r)^2 +\frac{ds}{dr}\int_{r}^{\infty}dt~[s(t)]\\
g&=&2:~~\nu(r) = s(2r)^2 -\frac{ds(2r)}{dr}\int_{0}^{2r}dt~[s(t)],
\eea
where the Fermi momentum $k_F$ is set equal to $\pi$ so that the maximum 
central density is unity. For all three forms of $\nu(r)$ given above, 
explicit calculations show that   
\beq
\int_{-\infty}^{\infty} \nu(x)dx = 
\int_{-\infty}^{\infty} [s(r)]^2dx = 1, 
\label{eq3}
\eeq 
independent of the value of $g$. Substituting this result in Eq.(\ref{eq2})
it follows that for interacting
bosons in CSM the fluctuation vanishes identically at $T=0$. 

While we cannot obtain the exact $\nu(r)$ in CSM for all $g$, the same may
be calculated for all values of $g$ in the harmonic lattice approximation.
The correlation function so obtained compares very well with the exact
correlation functions for $g=1/2,1,2$ and is given by \cite{diptiman}
\beq
\nu(x) ~=~\rho_0\sum_{n \ne 0} ~\Bigl( ~
\frac{1}{4\pi F(n,0)}~ \Bigr)^{1/2} ~\exp ~\Bigl[ ~- \frac{(x
\rho_0 -n)^2}{4F(n,0)} ~\Bigr] ~ -\rho_0,
\label{eq4}
\eeq
where $\rho_0=N/L$ is the average density and 
\beq
\lambda F(n,0) ~=~ \frac{1}{2\pi^2} ~\int_0^{\pi} ~dy ~\frac{1 - \cos 
~(yn)}{y~
-~ y^2/2\pi} ~=~ \frac{1}{2\pi^2} ~\int_0^{2 \pi} ~dy ~\frac{1 - \cos ~
(yn)}{y} ~.
\label{eq5}
\eeq
The above expression is given for completeness and its exact form is not
needed for further calculations. Again integrating over the real line we
get a result identical to that obtained using the exact correlation
functions in CSM. Thus the fluctuation vanishes identically for all $g$ in
this approximation at zero temperature. However, the result does not give
any indication of the behavior of the fluctuation at finite temperature.
To do this we take recourse to the mapping between the CSM and the
exclusion statistics first proposed by Haldane through a generalized Pauli
principle\cite{murthy,bernard}. 

A crucial property of exclusion statistical interactions is that they
should cause shifts in single particle energies at all scales\cite{ms}
(see next section).  This property is realized by a large class of one
dimensional models of interacting fermions where Fermi liquid theory
breaks down\cite{pwa,mss}. In fact it has been shown exactly that
quasiparticles with nontrivial exclusion statistics exist in a class of
models that are solved by the Bethe ansatz\cite{bernard}. Thus the result
obtained below should be, in principle, valid in a large class of models
with interactions. In particular, it is well known that the interacting
particle of CSM may be regarded as ideal exclusion statistics particles or
simply haldons\cite{murthy}.  The thermodynamic properties of an ideal 
gas of exclusion statistics particles have been investigated 
widely\cite{wilczek,isakov,yang,aoyama}.  
The distribution function has been computed and is given by
\beq
\langle n(\epsilon)\rangle~=~{1 \over (w(\epsilon)+g)},
\label{eq6}
\end{equation}
where $w(\epsilon)$ is the solution of the equation
\begin{equation}  
w(\epsilon)^g(1+w(\epsilon))^{(1-g)}~=~e^{\beta(\epsilon -\mu)} 
\label{eq7} 
\end{equation}
At zero temperature we have,
\beq
\langle n\rangle ~=~{1 \over g},~~~ for~~ \epsilon_k \leq  \mu
\label{eq61}
\end{equation}
and zero otherwise.

Note that the distribution function reduces to the usual Fermi and Bose
distribution functions for $g=1$ and $g=0$ respectively and in general $g$
is regarded as the exclusion statistics parameter. Indeed, the statistical
parameter $g$ is precisely the interaction strength as given in
Eq.~(\ref{ham}) in one dimension. However, the distribution function as
given above is valid in general and not necessarily restricted to one
dimensional models. The following discussion is therefore used for
illustration in the case of one-dimensional model but not restricted to
this case alone. 

The number fluctuation at a given energy $\epsilon_k$ is given 
by, 
\beq
(\delta n_k)^2 =   T \frac{\partial\langle  n_k \rangle}{\partial \mu}. 
\label{eq8}
\eeq 
Substituting for $\langle n_k\rangle$ from Eq.~(\ref{eq6}), we have
for total number fluctuation\cite{wilczek},
\bea
(\delta N)^2 &=& \sum_{k=0}^{\infty} w_k(1+w_k)\langle n_k \rangle^3 \\ ~\nonumber
           &=& \sum_{k=0}^{\infty} \langle  n_k \rangle (1-g\langle n_k 
\rangle) (1+(1-g)\langle n_k \rangle).
\label{eq9}
\eea 
The number fluctuation vanishes at $ T \rightarrow 0$ since $n_k
\rightarrow 1/g$ below the Fermi energy and zero otherwise. This result
holds no matter how weak the interaction strength is. However, at $g=0$, 
the bosonic limit, the number fluctuation diverges as noted earlier. 
In this exactly solvable model, we have thus shown that 
interactions do remove the fluctuation catastrophe encountered in the ideal 
Bose gas. 

While these results have been derived in one dimension, extension to
higher dimension is non-trivial since there is no suitable exactly
solvable many body model. However, it has been shown that models with
short range interactions in two dimension may be regarded as obeying
exclusion statistics in the mean-field picture\cite{sbm}.  We discuss the
fluctuation in these models in the next section. 

\subsection{Fluctuations in a two dimensional model}

We consider a two-dimensional system of bosons interacting via a zero-range 
repulsive pseudo-potential. The quantum dynamics is then
approximated  by the following Hamiltonian
\beq
H=\sum_{i=1}^N \left({p_i^2\over {2m}}+{1\over 2}m\omega^2 r_i^2\right)+
{2\pi\hbar^2\over m} \tilde {g} \sum_{i<j}^N \delta({\bf r}_i- {\bf r}_j)~,
\label{eq11}
\eeq
where the momenta and coordinates are planar vectors. The one-body
potential generated by the above zero-range interaction (including 
exchange) is 
\beq
U(n({\bf r}))={2\pi\hbar^2\over m }g~ n({\bf r}),~~~~~~  g=2 \tilde {g}~,
\eeq
where $n({\bf r})$ is the local number density of the system. 
In two-dimensions, $g\geq 0$ plays the role of the statistical parameter, 
with $g=0$ for non-interacting bosons. 
At finite temperature, for $T > T_c$, the Thomas-Fermi approximation 
yields\cite{bhaduri,brandon} 
\beq
n({\bf r})=
\int {d^2p/{(2\pi\hbar)^2}\over
{\left[\exp [({p^2\over {2 m}}+V(r)-\mu)\beta]-1\right]}}~,
\label{eq12}
\eeq
where the Thomas-Fermi mean potential $V(r)$ is given by
\beq
V({\bf r})=V_0(r)+{2\pi\hbar^2\over m} g~ n({\bf r})~.
\label{eq13}
\eeq
Here $V_0(r)$ is the one-body harmonic trap. Note that Eq.(\ref{eq12}) is 
valid only in the absence of a condensate. It has been 
shown~\cite{bhaduri,brandon}. however, that 
for a nonzero positive $g$, a self-consistent solution of this equation satisfying $\int n(r) d^2 r=N$ may be obtained right down to T=0. This solution 
has a lower free energy than the one with a condensate~\cite{mullin}, so 
we may take $T_c=0$ for $g>0$.   

 The momentum integration may be done 
analytically: 
\beq
n(r)= -\frac{m}{2\pi\hbar^2\beta}\ln\left[1-\exp [-\beta(V(r)-\mu)]
\right]~. 
\label{eq14}
\eeq
The local number fluctuation (between $r$ and $r+dr$) in GCE is given by
\beq
(\delta N)^2 = T \frac{\partial n(r)}{\partial \mu},
\label{eq15}
\eeq
where the coefficient of the temperature $T$ on the rhs is related to the 
compressibility. Taking the derivative of the local density with 
respect to the chemical potential we have
\beq
(\delta N)^2 = T\frac{m}{2\pi\hbar^2}~ \frac{1}{\exp [(V(r)-\mu)\beta]-1 +g }~,
\label{eq16}
\eeq
 
Note that $\mu$ is a function of temperature, and is determined by the 
condition that $\int n(r) d^2r=N$, and in the thermodynamic limit it 
approaches the lowest energy eigenstate as the temperature goes to zero. 
A few remarks on the thermodynamic limit are in order: The thermodynamic
limit is reached when $N\rightarrow \infty$ and $\omega\rightarrow 0$. 
However, in the limit of no confinement the density of states becomes a 
constant and there is no critical temperature below which condensation 
takes place. Preserving the density of states as in a two dimensional 
oscillator, the condensation temperature of an {\it ideal} bose gas 
is given by 
$$ T^{(0)}_c =(6/\pi^2)^{1/2}~ N^{1/2}\omega.$$
Thus the limit $N\rightarrow \infty$ and $\omega \rightarrow 0$ is 
obtained keeping $T_c^{(0)}$ constant.  
In an ideal bose gas ($g=0$), no self-consistent solution of Eq.~(\ref{eq14}) 
can be found for a fixed $N$ below this temperature $T^{(0)}_c$. 
However, when $g>0$  no matter how small, the self-consistent solution of 
Eq.~(\ref{eq14}) may then be found for all $T>0$. 

In this limit the fluctuation is given by,
\beq
(\delta N)^2 = T\frac{m}{2\pi\hbar^2} \frac{1}{\exp [({2\pi\hbar^2\over 
m} g n -\mu)\beta]-1 +g }~,
\label{eq17}
\eeq
where $n$ is the constant density given in the thermodynamic limit. 
(For the one-dimensional case, this was denoted by $\rho_0$ earlier.) 

In the absence of interaction, $g=0$, the chemical potential $\mu$ goes 
to zero at $T=T_c$ and the above expression diverges as expected. 
However, as in the one dimensional case, the  
fluctuation remains finite and approaches zero as $T\rightarrow 0$ when 
$g$ is finite, however weak the interaction may be.

The same result may be obtained in the non-interacting exclusion statistics 
description in two-dimensions. The local density as a function of the 
radial coordinate is given by \cite{bhaduri},
\beq
n({\bf r})=\int {d^2p/{(2\pi\hbar)^2}\over{\left[w + g\right]}}~,
\label{eq18}
\eeq
where the local variable $w(p,r)$ is defined through Wu's equation within 
TF approximation 
\beq
w^g (1+w)^{1-g} = \exp[\beta(\frac{p^2}{2m}+V_0(r)-\mu)],
\eeq
and $g$ is the exclusion statistics parameter which we will identify with
the interaction strength in the mean-field picture. We have retained the 
trap potential $V_0$ as in the interacting picture. 
Once again the momentum integration may be done easily and we obtain,
\beq
n(r)= \frac{m}{2\pi\hbar^2\beta}\ln\left[ \frac{1+w_0}{w_0}\right]~, 
\label{eq19}
\eeq
where the local variable $w_0(r)$ is determined through
\beq
w_0^g (1+w_0)^{1-g} = \exp[\beta(V_0(r)-\mu)].
\label{eq20}
\eeq
The fluctuation is then found by using Eq.~(\ref{eq15}) and we have
\bea
\delta N^2 &=& T \frac{\partial n(r)}{\partial \mu},\\
&=& T\frac{m}{2\pi\hbar^2} \frac{1}{w_0+g }~.
\label{katha}
\eea
In the bosonic limit taking $g=0$ in Eq.~(\ref{eq20}), we have $w_0 = 
\exp[\beta(V_0 -\mu)] - 1$. Substituting this in the above expression for 
the fluctuation in the thermodynamic limit it is easy to see that at 
$T_c$ there is divergence. However, for positive definite $g$ there is no 
divergence. Furthermore the equivalence between the non-interacting 
exclusion statistics picture and the mean-field description is 
established using the following relationship
\beq
w_0 (r) = \exp[\beta(V(r)-\mu)] -1,
\label{kahini}
\eeq
where $V(r)$ is the self consistent mean-field potential. Substituting this 
in Eq.(\ref{katha}), and taking the thermodynamic limit yields Eq.(\ref{eq17}).
Note that this 
equivalence allows one to calculate fluctuations in either the mean field 
picture or in the non-interacting exclusion statistics picture. 
Indeed this holds for the computation of other global thermodynamic 
quantities as well.

\section{ Ground state fluctuation in the canonical ensemble}

The total number fluctuation defined in the previous section in the GCE is
obtained by summing over all the single particle states. In the canonical
ensemble however, the number of particles is fixed and therefore the
fluctuation in particle number has to be defined with respect to a reference
state. One way defining the fluctuation is to look at the ground state
occupancy as a function of temperature, which is present even when the
total number of particles in the trap is fixed. At $T=0$ all the particles
are in the ground state. At a nonzero temperature (or excitation energy), 
there are many ways of exciting the particles from the ground state, leading to
a fluctuation in the ground state population.  This number fluctuation for
non-interacting bosons in a harmonic trap has been calculated by a number
of authors~\cite{grossman} as a function of the excitation energy or
temperature.  This has also been calculated for fermions in the CE by us
in a previous publication\cite{tran}. Unlike the case of GCE for bosons,
the ground state fluctuation in the CE is finite at all temperatures.

We may extend the analysis of fluctuations in CE to particles interacting
via the inverse square pair-wise interaction in one dimension. It is more  
convenient to perform the calculation in the fermionic basis, although 
the formulae given here are applicable for both interacting fermions, or 
interacting bosons. In this section, quantities like energy and  
ground state number fluctuation of the interacting system with 
interaction strength 
$g$ will be denoted by a bracketed superscript, e.g. $E^{(g)}$, 
$((\delta N_0)^2)^{(g)}$. 
We recall that the spectrum of the CSM Hamiltonian given in 
Eq.~(\ref{ham}) is exactly known.  The states may be labeled 
by a set of fermionic occupation numbers $\{n_k\},~ k=1,...,\infty,~ n_k =
0,1$. The energy $E^{(g)}$ of the system in the fermionic basis is then 
given by,
\begin{equation}
E^{(g)}\{n_k\} = \sum_{k=1}^{\infty} \epsilon_k n_k - \omega  (1-g)   
\frac{N(N-1)}{2},
\label{energy}
\end{equation}
where $\epsilon_k = (k-\frac{1}{2}\hbar\omega)$ denotes the harmonic 
oscillator energy levels and $ N=\sum_{k=1}^\infty n_k$.  
As can be seen
from Eq.~(\ref{energy}), the effect of the interaction is that each
particle shifts the energy of every other particle by a constant
$\hbar\omega (g-1)$.  The energy functional can also be written as
\begin{equation}
E^{(g)}\{n_k\} = \sum_{k=1}^{\infty} \epsilon_k n_k - \omega  (1-g)
\sum_{k_1<k_2=1}^{\infty} n_{k_1} n_{k_2}.
\label{enfun}
\end{equation}
The exact spectrum of the model is thus reproduced by an effective
Hamiltonian of quasi-particles with constant density of states and
constant Landau parameters. As mentioned before, this scale invariant
energy shift is the basic reason for the occurrence of nontrivial
exclusion statistics where $g$ plays the role of exclusion statistics
parameter with $g=0,1$ for bosons and fermions.

The general  canonical partition function in any basis is  written in the 
occupation number representation as\cite{murthy}
\begin{equation}
Z_N^{(g)} = \sum_{\{n_k\}} e^{-\beta E^{(g)}\{n_k\}}.
\label{occnumrep}
\end{equation}
Using the energy spectrum in CSM given in Eq.~(\ref{energy}), the 
N-particle partition function in this one dimensional model is given by  
\begin{equation}
Z_N^{(g)} = e^{\tilde \beta (1-g) \frac{N(N-1)}{2}} Z_N^F,
\label{zna} 
\end{equation}  
where $\tilde \beta = \beta \hbar \omega$ and $Z_N^F$ is the N particle
fermion partition function.  Setting $g=0$, the bosonic partition
function is obtained,
\begin{equation}
Z_N^B = e^{\tilde \beta \frac{N(N-1)}{2}} Z_N^F.
\label{znb}
\end{equation}
Combining Eqs.~(\ref{zna}) and (\ref{znb}) we may write the partition
function for CSM as, 
\begin{equation}
Z_N^{(g)} = (Z_N^F)^g (Z_N^B)^{1-g}.
\label{znag}
\end{equation}
The canonical partition function given above is exact in CE and may be used 
for calculating the thermodynamic properties of the system in CSM within 
the canonical ensemble formalism. 
The moments of the occupation number are related to the partition function by
\bea
\langle n_k \rangle^{(g)} &=&\frac{1}{Z_N^{(g)}}y_k 
\frac{\partial Z_N^{(g)}}{\partial 
y_k},\\
\langle n_k^2 \rangle^{(g)} &=&\frac{1}{Z_N^{(g)}}y_k \frac{\partial}{\partial 
y_k} \left(y_k\frac{\partial Z_N^{(g)}}{\partial y_k}\right),
\label{occuag}
\eea
where $y_k = \exp(-\beta\epsilon_k)$. Therefore, it follows that
\beq
\langle(\delta n_k)^2\rangle^{(g)}=y_k \frac{\partial \langle 
n_k\rangle^{(g)}}{\partial y_k}. 
\eeq
Using Eqs.~(\ref{znag})-(\ref{occuag}), $\langle n_k \rangle^(g)$ can be expressed in terms of those of fermions and bosons via: 
\begin{equation}
\langle n_k \rangle^{(g)}=g\langle n_k \rangle^F+(1-g)\langle n_k \rangle^B 
\label{occuag0}
\end{equation}
Unlike $\langle n_k \rangle^{(g)}$, $\langle n_k^2 \rangle^{(g)}$ does not have a simple form as Eq.~(\ref{occuag0}).  However, the expression for the fluctuation in the occupation number does:
\beq
((\delta n_k)^2)^(g) = g((\delta n_k)^2)^F + (1-g)((\delta n_k)^2)^B.
\label{fluctgoccuag}
\eeq
Eq.~(\ref{fluctgoccuag}) gives only the fluctuation in the occupation of a given level $k$, while the quantity we are seeking is the {\it ground state} number fluctuation.  The latter is formally defined in any ensemble as:  
\beq
(\delta N_0)^2 = \sum_{k}(\delta n_k)^2=\sum_{k} (\langle n_k^2\rangle - \langle n_k \rangle^2) 
\label{fluctgs}
\eeq
where the sum $k$ runs over only the levels which are completely
occupied at zero temperature. Thus, in an ab initio calculation, one would formally sum over the quasiparticle levels which are occupied at $T=0$ to get $((\delta N_0)^2)^(g)$.  Fig.~1 shows the level flow in CSM as a function of $g$ obtained from Eq.~(\ref{enfun}) at $T=0$.  It can be seen that as $g$ changes from the fermionic to the bosonic end, the number of levels contributing to the ground state remains constant, while the Fermi energy decreases accordingly.  This means that one may obtain $((\delta N_0)^2)^{(g)}$ by simply substituting the ground state fluctuations for fermions and bosons. {\it ie:}    
\beq
((\delta N_0)^2)^{(g)} = g((\delta N_0)^2)^F + (1-g)((\delta N_0)^2)^B
\label{fluctg}
\eeq

\section{Discussions}

In the previous section we have discussed the ground state number fluctuation in GCE and CE in an interacting system.  We now compare the results obtained in these two ensembles.  Due to numerical difficulty at low temperatures, we are able to do the comparison only for $N=10$ particles.  Despite this limitation, some interesting points can be made.  In Fig.~2(a), we show the behavior of the relative ground state fluctuation against temperature of interacting bosons for both ensembles for interacting strengths of $g=0,1/2,1$. The $g=0$ case corresponds to free bosons.  Fig.~2(b) shows the low temperature region of the curves in Fig.~2(a).  Note that as $T \rightarrow 0$ the GCE fluctuation for free bosons diverges as expected, whereas those for interacting bosons remain finite and approach zero.  Note also that in the ideal  Haldane gas picture the $g=1/2$ case corresponds to semions and the $g=1$ case is the non-interacting fermionic limit. The GCE fluctuation for semions is found using Eqs.~(\ref{eq6}), (\ref{eq7}) and (\ref{eq9}), where $\mu$ is determined by the constraint that the average total number of particles is $N$. 

Unlike in GCE, the CE ground state fluctuation for free bosons remain finite as $T \rightarrow 0$. The ground state fluctuation of free bosons, however, approaches zero {\em exponentially} (see Fig.~2(b)), contrary to previous results that find a linear dependence with $T$ all the way to $T=0$ \cite{grossman,thesis}. In Appendix A we give the low temperature expansions of the fluctuation squared $(\delta N_0)^2$ in powers of $x$, where $x=e^{-\tilde \beta}$.  Clearly, at very low temperatures $(\delta N_0)^2$ is independent of the number of particles $N$ (in both GCE and CE).  Therefore, the exponential behaviour of the fluctuation of free bosons at low temperatures should  remain valid even in the large $N$ limit.  In the case of fermions the CE and GCE fluctuations are similar except at very low temperatures. Clearly, the CE and GCE curves of fermions approach zero differently as $T\rightarrow 0$. For high temperatures, on the other hand, GCE and CE give identical results, as expected.   

It is to be noted from Fig. 2 (a) that the number fluctuation for  
free bosons and free fermions cross at a certain temperature, with the  
fermion one getting bigger at higher temperatures. This is to be expected, 
since the number of possibilities of creating holes within the fermi sea, 
and distributing particles above, increases more rapidly than for bosons, 
whose ground state has only one level. Semions ($g=1/2$), which may be 
regarded either as interacting bosons, or as interacting fermions, have 
fluctuations intermediate between those of bosons ($g=0$) and 
fermions ($g=1$).

In one dimension, we have calculated the number fluctuation of bosons interacting via CSM using the ideal Haldane picture for both GCE and CE.  The next challenge is to find the {\em exact} fluctuation using combinatorics.  This has been done for free fermions which correspond to interacting bosons with interacting strength of $g=1$ \cite{tran}.  Work is in progress to calculate this exact fluctuation for a general $g$. 

%As in the case of non-interacting fermions and bosons, one may also calculate the exact ground state fluctuation for ideal haldons using counting rules\cite{count} in the microcanonical ensemble (MCE). This proves to be non-trivial, however, since the energy levels in the ideal Haldane picture get shifted depending on the occupation of the levels below as can be seen from Eq.~(\ref{enfun}).  If this can be done, one would expect a similar relationship to Eqs.~(\ref{eq9}) and (\ref{fluctgs}) to exist in the microcanonical ensemble. 

\acknowledgements {Two of us (RKB and MNT) thank the Institute of 
Mathematical Sciences, where part of this work was done, for hospitality. 
This work was supported by Natural Sciences and Engineering Research 
Council (NSERC) of Canada. }

\section*{Appendix: Low temperature expansions in GCE and CE}

The low temperature behaviour of thermodynamic quantities in GCE are well
known for bosonic systems. However, a comparison between GCE and CE
calculations at low temperatures is not usually discussed for either Bose
or Fermi systems.  Further, making use of some asymptotic expansions, the CE 
fluctuation for bosons was earlier found to be linear right down to $T=0$.  However, we give here the expansion of the fluctuation squared at low temperature in power of $x$, where $x=e^{\tilde \beta}$, and show that the CE 
fluctuation of bosons is in fact exponential at very low $T$.  In GCE only expansion for
fermions is possible, since the fluctuation tends to infinity at low
temperature for bosons. Both expansions are possible in CE. 

\paragraph{Grand Canonical Ensemble:}

In GCE the (fermionic) occupation number is:
\beq
\langle n_k \rangle_{GCE}= \frac{1}{x^{(\mu -\epsilon_k)}+1}= 
\frac{1}{x^{(\mu -k-1/2)}+1} 
\eeq
for a one dimensional system. The ground state number fluctuation squared 
is given by: 
\beq
\langle (\delta N_0)^2 \rangle_{GCE}=\sum_{k=0}^{k_F} 
\frac{x^{(\mu-k-1/2)}}{\left[x^{(mu -k-1/2)}+1\right]^2} 
\eeq
where $k_F$ is the Fermi level.  At low temperatures, for the one-dimensional 
harmonic oscillator, $\mu \approx \mu_0 
=N$.  Therefore, 
\beq
\langle (\delta N_0)^2 \rangle_{GCE}=\sqrt 
x-2x+4x^{3/2}-4x^2+6x^{5/2}-8x^3+8x^{7/2}-8x^4...~~ , ~~ N \geq 4. 
\eeq
Note that the first few terms are independent of $N$.

\paragraph{Canonical Ensemble:}

In CE the first and second moments of the occupation number are
known~\cite{parvan}: 
\begin{eqnarray}
\langle n_k \rangle &=& 
\frac{1}{Z_N}\sum_j^N(\pm)^{j+1}x^{j\epsilon_k}Z_{N-j}, \\ 
\langle n_k^2 \rangle &=&\frac{1}{Z_N}\sum_j^N(\pm)^{j+1}\left[j \pm 
(j-1)\right]x^{j\epsilon_k}Z_{N-j},  
\end{eqnarray}
where the upper and lower signs refer to bosons and fermions 
respectively.  Summing over the ground states up the Fermi level gives the 
fermionic ground state number:  
\beq
\langle N_0 \rangle_{CE} = 
\frac{1}{Z_N}\sum_j^N(-1)^{j+1}x^{j/2}\frac{1-x^{jN}}{1-x^j}~Z_{N-j}, 
\eeq
where we let $\epsilon_k=k-1/2$, with $k=1,2,...$. Therefore,
\begin{eqnarray}
\langle (\delta N_0)^2 \rangle_{CE} &=& \sum_{k=1}^N\langle n_k 
\rangle-\sum_{k=1}^N\langle n_k \rangle^2 \nonumber \\  
%&=&\frac{1}{Z_N}\sum_j^N(-1)^{j+1}Z_{N-j}~x^{j/2}
%\left[\frac{1-x^{jN}}{1-x^j}Z_{N-j}-\frac{1}{Z_N}
%\sum_k^N(-1)^{k+1}Z_{N-k}~x^{k/2}
%\frac{~1-x^{(j+k)N}}{1-x^{(j+k)}}\right], \\ \nonumber        
&=&x + 2x^4+... ~~ , ~~ N\geq 4,
\end{eqnarray}
where we have used $\langle n_k \rangle=\langle n_k^2 \rangle$ and again 
the first few terms are independent of the system size $N$.

For bosons the ground state consists of one single lowest level, the 
low temperature expansion of the number fluctuation is given by 
\begin{eqnarray}
\langle (\delta N_0)^2 \rangle_{CE} &=& \langle n_0^2 \rangle - 
\langle n_0\rangle^2\nonumber\\ 
%\sum_{k,k'}^N\mathop{\prod_{i=N-k+1}^N}_{i'=N-k'+1}(1-x^i)(1-x^{i'}) \\ 
%\nonumber  
&=& x+3x^2+4x^3+7x^4+... ~~,~~ N>4.
\end{eqnarray}
Again as in the fermionic case the first few terms in the low temperature expansions are 
independent of the system size. Indeed it is interesting to note that in 
CE, the fluctuations in both the systems approach zero as $T\rightarrow 
0$ in exactly identical fashion.

%~
%\begin{figure}[htp]
%\vskip 20truecm
%{\special{psfile="/home/tran/CHENNAI/fig1.eps"
%          hscale=90 vscale=90 hoffset=-50 voffset=-40}}

%\caption{The level flow of quasi-particle energy levels in CSM as a 
%function of $g$.} 
%\end{figure}

%\newpage
%~
%\begin{figure}[htp]
%\vskip 20truecm
%{\special{psfile="/home/tran/CHENNAI/fig2a.eps"
%          hscale=90 vscale=90 hoffset=-45 voffset=-40}}

%\caption{(a) The ground state fluctuation in GCE and CE as a function of 
%temperature for $N=10$. We show the results for fermions, bosons and also semi%ons ($g=1/2$).}  
%\end{figure}

%\setcounter{figure}{1}
%\newpage
%~
%\begin{figure}[htp]
%\vskip 20truecm
%{\special{psfile="/home/tran/CHENNAI/fig2b.eps"
%          hscale=90 vscale=90 hoffset=-30 voffset=-40}}

%\caption{(b) Same as in (a) but using the low temperature expansions 
%given in the Appendix}.    
%\end{figure}

\newpage
\begin{figure} 
\label{figure1}
\caption{The level flow of quasi-particle energy levels in CSM as a 
function of $g$.}
\vspace{7 mm}
\label{figure2(a)}
\caption{(a) The ground state fluctuation for the one-dimensional CSM system 
in GCE $(Eq.~(\ref{eq9})$  and CE $(Eq.~(\ref{fluctg})$ as a function of 
temperature for $N=10$. We show the results for fermions, bosons and also semions ($g=1/2$).}  
\vspace{7 mm}
\label{figure2(b)}
\caption{(b) Same as in (a) but using the low temperature expansions 
given in the Appendix}.    
\end{figure}
\newpage

\end{document}